\documentclass{article}
\usepackage{amsmath}
\usepackage{icrctc07}

\title{Search for correlation of UHECRs and BL Lacs \\
 in Pierre Auger Observatory data}
\shorttitle{Search for correlation with BL Lacs}
\authors{Diego Harari$^1$, for the Pierre Auger Collaboration$^2$}
\shortauthors{D. Harari, for the Pierre Auger Collaboration}
\afiliations{$^1$Departamento de F\'\i sica, Centro At\'omico Bariloche, 
CNEA and CONICET, Argentina\\
$^2$Observatorio Pierre Auger, Av. San Mart\'{\i}n Norte 304, 
(5613)
  Malarg\"ue, Mendoza, Argentina}
\email{harari@cab.cnea.gov.ar}

\abstract{Several analyses of the data collected by other experiments have found an 
excess of cosmic rays in correlation with subclasses of BL Lacs.
Data from the Pierre Auger Observatory do not support previously reported excesses.  
The number of events correlated with BL Lac positions is compatible with that expected 
for an isotropic flux.}

\begin{document}

\maketitle

\section{Introduction}

Anisotropies in the arrival directions of UHECRs are likely to provide significant 
clues to their origin and nature. One potential type of sources of UHECRs are BL Lacertae 
objects, which are a subclass of blazars, active galaxies with beamed emission from a
relativistic jet which is aligned roughly toward our line of sight.   
A correlation larger than that expected on average for an isotropic flux was found \cite{TT01} 
between a subset of BL Lac positions and arrival directions of UHECRs recorded by the 
Akeno Giant Air Shower Array (AGASA) with energies above 48~EeV 
and by the Yakutsk experiment with energies above 24~EeV (1~EeV $= 10^{18}$~eV). 
This correlation, as well as others reported in \cite{TT02,GTTT02}
between BL Lacs and UHECRs recorded by the AGASA and Yakutsk experiments, 
were not supported by the data collected by the 
High Resolution Fly's Eye (HiRes) air fluorescence 
detector \cite{hires}. However, HiRes registered an excess of 
correlations with a subset of BL Lacs 
on a scale consistent with its angular resolution, with maximum
significance for events with energies above 10~EeV \cite{hires,GTTT04}.
This excess occurs in an energy range and angular scale different from 
previously reported correlations. An assessment of its statistical 
significance requires the analysis of independent data.  

We use data recorded by the Surface Detector of the Pierre Auger 
Observatory between 1 January 2004 and 15 March 
2007 to search for cross-correlations with BL Lacs,
particularly to test previous potential signals. 
The number of events with energies above
10~EeV in the present analysis is more than 6 
times larger than  used in preceding cross-correlation searches.
Our data do not support previously reported excesses. 

\section{Data set and methods}\label{dataset}

Hybrid operation of the Pierre Auger Observatory  
allows precision energy calibration
of the large number of events recorded by its Surface Detector (SD),
as well as several consistency checks to be performed.
The energy and angular reconstruction accuracy of the SD 
are described in detail elsewhere \cite{ave}.
The quality trigger implemented in the present analysis requires at 
least five surface stations around 
that with the highest
signal to be active when the event was recorded, and that the reconstructed shower core be inside a triangle of active
stations \cite{allard}. 
We use events recorded by the SD with energies above 3~EeV and zenith angles 
smaller than $60^\circ$.
There are 14143 events in the data set, of which 1672 have energies above 
10~EeV.
This set does not include a small 
fraction of events with energies above 10~EeV that triggered less than 6 surface stations,
nor events with energies below 10~EeV that triggered less than 4 stations. 
The 
angular 
resolution
of the SD array is defined as the angular aperture around an arrival direction of 
CRs within which  68\% of the showers are reconstructed.
It is $0.9^\circ$ for events with energies above 10~EeV and 6 or more stations 
triggered, and
$1.2^\circ$ for events with energies above 3~EeV and 4 or more stations triggered.

The acceptance area of the SD is saturated for events with energies above
3~EeV, and is only limited by geometric effects, which produces a simple
analytic dependence upon declination.  The small modulation of the
exposure in right ascension originated by the present continuous array
growth as well as from detectors dead periods can be estimated from the
number of active stations as a function of time. It can be ignored, since
it has negligible effects upon the analysis performed in this work. It is
then straightforward to evaluate the probability $p$ that an individual
event from an isotropic flux has its arrival direction less than a given
angular distance away from at least one member of a collection of
candidate point sources.  The probability that $k$ or more out of a total
of $N$ events from an isotropic flux  are correlated by chance is
given by the cumulative binomial distribution $P=\sum_{j=k}^N
\left(\begin{matrix}{N}\cr{j}\end{matrix}\right)  p^j(1-p)^{N-j}$.  The
significance of $P$ is controversial \cite{EFS} if the parameters of the
search, such as the angular scale, energy threshold and the collection of
candidate sources, are not fixed a priori. An estimate of the chance
probability for a particular correlation search is given by the fraction
of simulated isotropic sets that have a smaller or equal value of $P$ than
the data anywhere in the parameter space, after a scan in the angular
scale and energy threshold \cite{FW}.

\section{Test of previous correlation signals}

We test previously reported correlations between UHECRs and subsets of BL Lacs.
Note that we test the physical hypothesis of correlation with a particular class of objects
at a given angular scale and above a given energy threshold, but 
the collections of candidate sources  are not identical to those in the original reports,
because the sky observed by the southern Pierre Auger Observatory is different, and has only a partial
overlap.

\begin{itemize}
\item {\bf Test A:}
22 BL Lacs from the $9^{\rm th}$ edition of the catalog of quasars and active nuclei \cite{VC06}, 
with optical magnitude $m < 18$, redshift $z > 0.1$ or unknown, 
and 6~cm radio flux $F_6>0.17$~Jy. 8 of these BL Lacs are in the field of view (f.o.v.) of the Pierre Auger Observatory
with zenith angles smaller than $60^\circ$. 
\item {\bf Test B:}  
157 BL Lacs (76 in the f.o.v.) from the $10^{\rm th}$ edition of \cite{VC06}  
with  $m<18$.
\item {\bf Test C:} 
14 BL Lacs (3 in the f.o.v.)
selected on the basis of possible association with $\gamma$-ray sources \cite{GTTT02}.
\item {\bf Test D:} 204 confirmed BL Lacs (106 in the f.o.v.) from the $10^{\rm th}$ edition of \cite{VC06} with $m < 18$. 
Subclasses: a) 157  BL,
b) 47 HP. 
\end{itemize}

Confirmed BL Lacs are classified by spectral properties as BL or HP (high 
optical polarization) in \cite{VC06}.
The objects tested in cases A, B and C are those classified as BL only. 

Table~1 summarizes the results of the tests. 
It lists the case considered, the reference for the original report, 
the lower energy threshold, the number of events with energy above that threshold, the angular size of the
search, the number of events observed to correlate within that angular 
size, the mean number of correlations 
expected by chance for an isotropic flux, and the chance probability $P$. 
 
  \begin{table*}
    \begin{center}
      \begin{tabular}{|c|c|c|c|c|c|c|c|c|}
        \hline
        Test & Ref. & $E_{th}$ & Number of  & Angular size &
        Observed & Expected & Probability \\ 
         &  & (EeV) & events &  &
         & (isotropic) &  \\ 
\hline
A & \cite{TT01} & 24  &  267 & $2.5^{\circ}$ &  1  &  1.0 &  0.63 \\      
        \hline
B & \cite{TT02} & 40  &  62 & $2.5^{\circ}$ &  2  &  2.5 &  0.71 \\      
        \hline
C & \cite{GTTT02} & 24  &  267 & $2.9^{\circ}$ &  1  &  0.5  &  0.41 \\      
        \hline
D & \cite{hires} &   &   &  &  11 &  12.1  &  0.66 \\      
        
 \quad a) & \cite{hires,GTTT04} & 10  & 1672   & $0.9^{\circ}$ &  8  &  
8.9  &  0.67 \\      
        
 \quad b) & \cite{hires} &  &  &  &  3  &  3.2  &  0.62 \\      
        \hline
      \end{tabular}
      \caption{Summary of tests of previously reported correlations. 
See the text above for details.
}\label{tests}
    \end{center}
  \end{table*}

Our data do not support any of the previously reported correlation excesses.
There is no significant correlation either, when the tests are performed
with the same selection criteria 
against the BL Lacs in the latest ($12^{\rm th}$) 
edition of the catalog of quasars and active nuclei \cite{VC06}. Nor is there any significant excess if the lower energy 
thresholds are changed $\pm 20\%$ from those of preceding analyses, to account for potential differences 
in energy calibration between different experiments.

The determination of the statistical significance with which our measurements exclude the hypothesis that the signal present in the HiRes data set (case D) is due to correlations with BL Lacs is a delicate issue. The sky observed by the two experiments is not the same. Catalog incompleteness and the possibility of different selection effects in the two fields of view additionally complicate comparisons. The HiRes data set has 271 events with energies above 10~EeV. Its correlation signal is best fit, using a maximum likelihood method \cite{hires}, with $n_s=11$ cosmic rays that come from source positions ($n_s=8$ from objects classified as BL and $n_s=3$ from objects classified as HP). Our correlation search was performed 
at the scale of the angular resolution of the SD array ($0.9^\circ$). We thus expect to 
reconstruct inside a search window 68\% of the showers initiated by CRs with arrival 
direction coincident with a BL Lac position.
There are 106 confirmed BL Lacs with $m<18$ in the field of view of the Pierre Auger Observatory, and 186 in the HiRes case. The ratio between the number of candidate sources in each field of view, weighted by the respective relative exposure, is approximately 0.4. 
Assuming that the degree of correlation is comparable in different portions of the 
sky, the excess correlation in HiRes data suggests 
the hypothesis that, using the same catalog and selection criteria, the correlation
between BL Lac positions and UHECRs arrival directions should be of greater statistical significance in the Pierre Auger Observatory
data set. Normalization to the signal in HiRes data suggests an expectation of  
$11\times (1672/271) \times 0.68 \times 0.4\approx 18.5$ events within $0.9^\circ$ from 
candidate sources in the Pierre Auger Observatory field of view, in 
addition to a mean of 12.1 events from an isotropic background.
The observation of a total of 11 events is strongly against the correlation hypothesis.

\section{Extended search}

We have extended our search for correlations with BL Lac positions to 
energy ranges and angular separations different than those that gave 
maximum signals in previous analyses. These extended searches also serve 
to account for potential differences in energy calibration and angular 
accuracy between different experiments that could make a possible 
correlation signal appear in a different range of parameters.

\begin{figure*}
\begin{center}
\noindent
\includegraphics [width=0.34\textwidth,angle=-90]{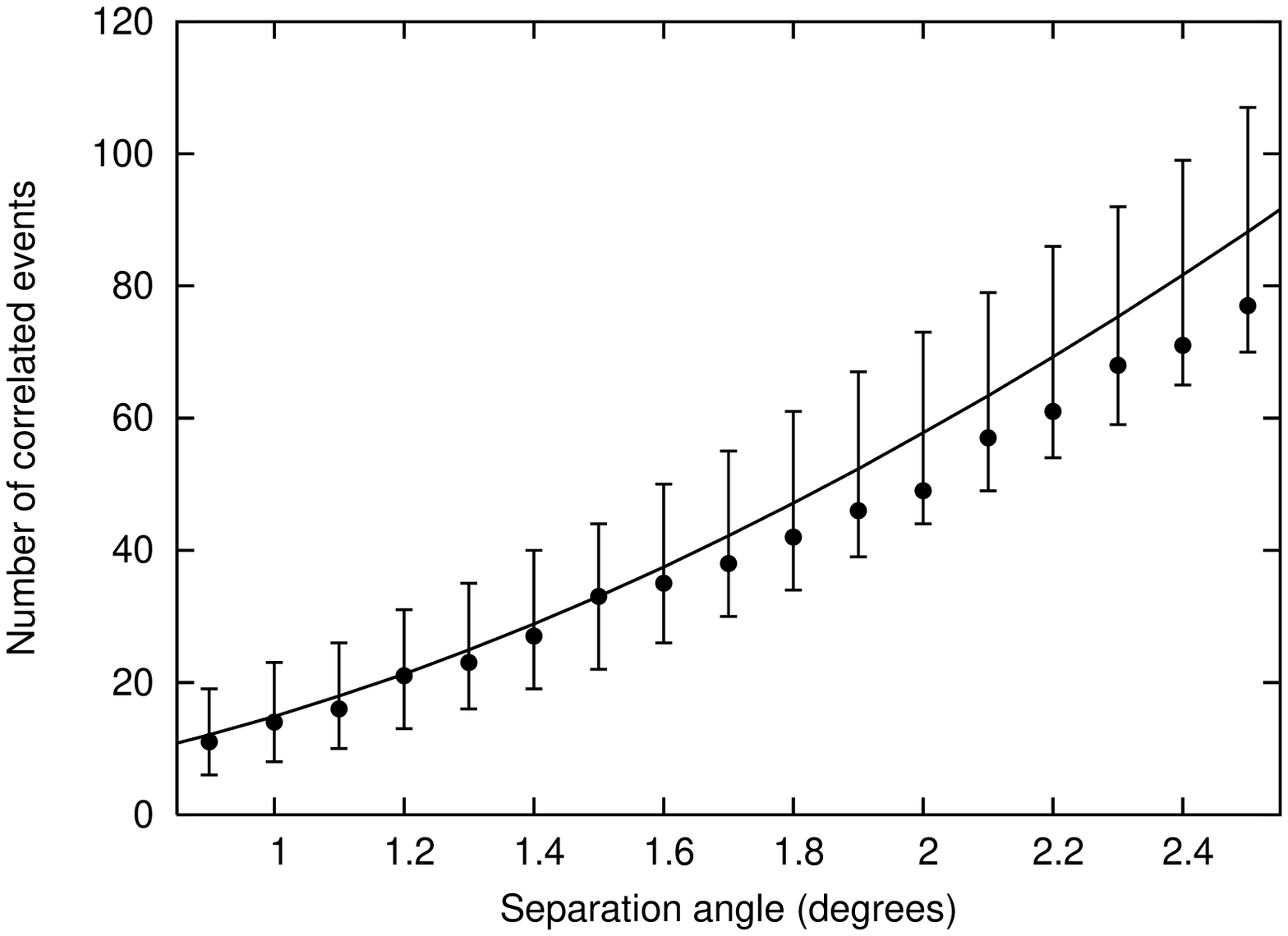}
\includegraphics [width=0.34\textwidth,angle=-90]{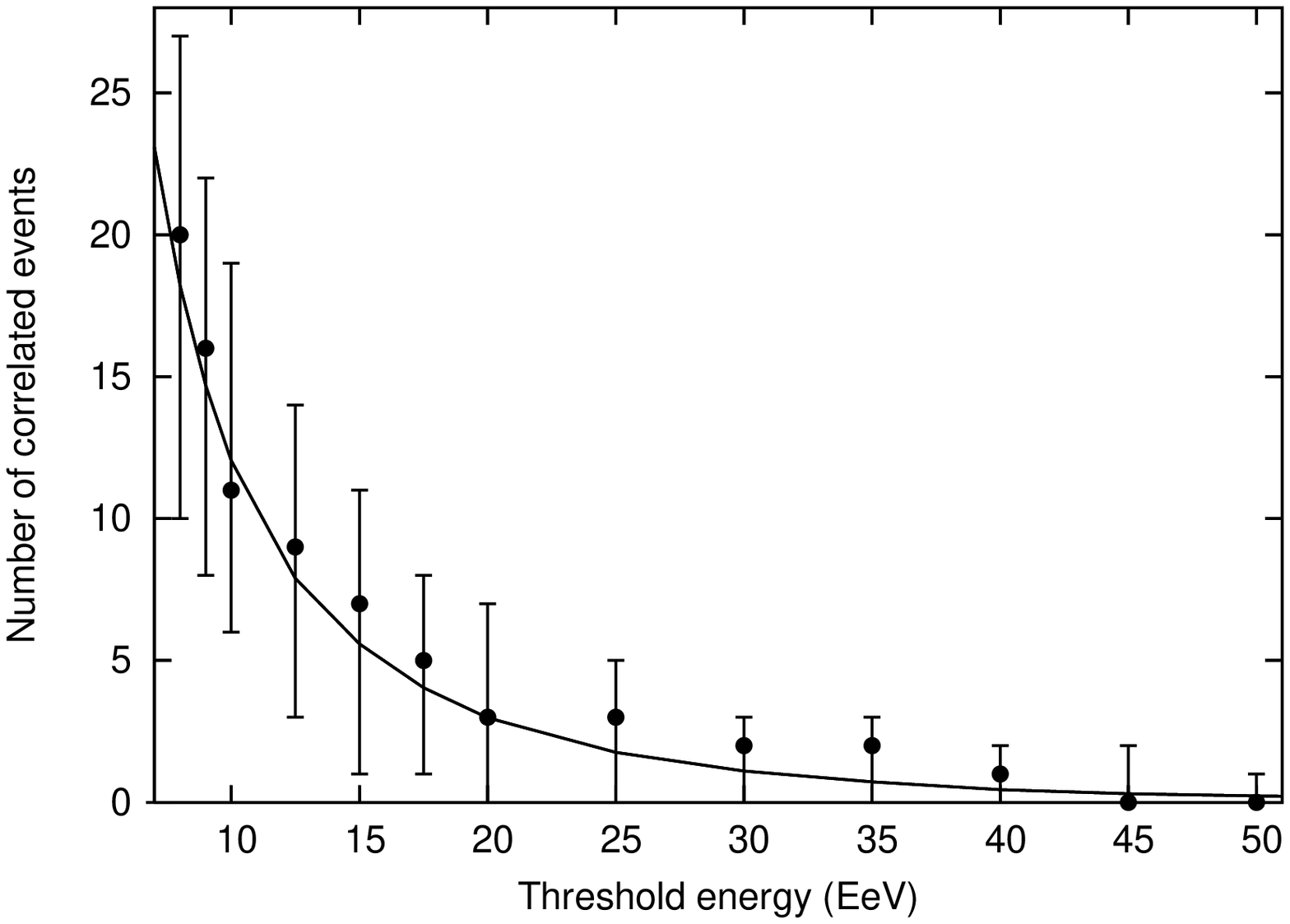}
\end{center}
\caption{Number of events correlated with confirmed BL Lacs with 
optical magnitude $m < 18$ from the 
$10^{\rm th}$ edition of the catalog of quasars and active galactic nuclei 
\cite{VC06}
 (points) and average for an isotropic 
flux 
(solid line) along with dispersion in 95\% of simulated isotropic sets (bars). Left: as a function of the angular separation (threshold energy fixed at 10~EeV). Right: as a function of threshold energy (angular separations below $0.9^\circ$).}\label{fig1}
\end{figure*}

As an illustration, we plot in the left panel of Figure~1 the number of 
CRs with energies above 10~EeV that are correlated with any of the 204 confirmed BL Lac positions of case D
as a function of angular distance. The solid line is  the mean number of correlations expected by chance for an
isotropic flux. Fluctuations in 95\% of simulated isotropic sets are contained 
within the bars. The right panel is the analogous plot as a function of threshold energy, for angular separations below $0.9^\circ$. There is not an excess for the specific energy threshold and angular separation tested in the section above, nor is any significant excess found for neighboring values of those parameters.  

We have extended the search for correlations with all the subclasses of BL Lacs in the previous section, selected both from
the catalog versions used in preceding searches as well as from its latest (12$^{\rm th}$) edition. 
We have scanned the lower energy threshold starting from 3~EeV, eliminating the event with lowest energy in each scan step. We have scanned the angular separation starting from $0.9^\circ$ for energies above 10~EeV and from $1.2^\circ$ for lower energy thresholds
(the decrease in angular resolution does not justify a scan at smaller separations). The angular separation was scanned
up to $3^\circ$. 
The search gave no significant correlation excess. The smallest value found for the 
probability $P$ that the observed correlation in a given scan step 
happened by chance under isotropic conditions was $P=0.03$. This value 
corresponds to the observation of 6 CRs among the subset of the 69 events
with energy above 38.8~EeV with arrival direction less than 
$2^\circ$ away from one of the 204 confirmed BL Lacs of case D (4 from 
objects classified BL, 2 from objects classified HP), while 2.4 are 
expected on average for an isotropic flux (1.8 around BL objects, 0.6 
around HP BL Lacs). 
Since 12\% of simulated isotropic sets have equal or 
smaller value of $P$ somewhere in the parameter space after a similar scan,
the excess observed is compatible with expected fluctuations under isotropic conditions.

\section{Conclusion}\label{conclusion}

Data from the Pierre Auger Observatory, with 6 times more events with energy above 
10~EeV than used in preceding searches, do 
not support previously reported excesses of correlation between the arrival
directions of UHECRs and subclasses of BL Lacs. 
The number of correlations found is compatible with that expected 
for an isotropic flux.

\bibliography{icrc0075}
\bibliographystyle{unsrt}

\end{document}